\documentclass[12pt]{article}
\usepackage{amsmath} 
\usepackage{longtable}
\usepackage{graphicx,epsfig}    
\usepackage{color}
\usepackage{cite}
\usepackage{amssymb}
\newcommand\as{\alpha_{\mathrm{S}}} 
\newcommand\f[2]{\frac{#1}{#2}}

\def\dPS{d\text{PS}_3} 
\def\beq{\begin{equation}} 
\def\eeq{\end{equation}} 
 
\def\nn{\nonumber}

\def\b0{\beta_0}

\def\beeq{\begin{eqnarray}}
\def\eeeq{\end{eqnarray}}
\def\ep{\epsilon}
\def\bom#1{{\mbox{\boldmath $#1$}}}

\def\mur{\mu_R} 
\def\muf{\mu_F}
\def\mur2{\mu_R^2} 
\def\muf2{\mu_F^2}

\newcommand {\apgt} {\ {\raise-.5ex\hbox{$\buildrel>\over\sim$}}\ }
\newcommand {\aplt} {\ {\raise-.5ex\hbox{$\buildrel<\over\sim$}}\ }

\begin{document} 

\setlength{\parskip}{0.15cm}
\setlength{\baselineskip}{0.52cm}

\begin{titlepage}

\begin{flushright}
ICAS 22/16
\\
ZU-TH 37/16
\end{flushright}

\renewcommand{\thefootnote}{\fnsymbol{footnote}}
\thispagestyle{empty}
\noindent
\vspace{0.5cm}

\begin{center}
{\bf \Large 
Two-loop corrections to the triple Higgs boson
\\
  production cross section \\}
  \vspace{1.25cm}
{\large
Daniel de Florian$\,^{(a)}$\footnote{deflo@unsam.edu.ar} and 
Javier Mazzitelli$\,^{(a,b)}$\footnote{jmazzi@physik.uzh.ch} \\
}
 \vspace{1.25cm}
 {
    $^{(a)}$ International Center for Advanced Studies (ICAS), ECyT-UNSAM, Campus Miguelete \\
25 de Mayo y Francia, (1650) Buenos Aires, Argentina\\[0.3cm]
$^{(b)}$ Physik-Institut, Universit\"at Z\"urich, \\ 
Winterthurerstrasse 190, CH-8057 Z\"urich, Switzerland
\\
 }
  \vspace{1.5cm}
  \large {\bf Abstract}
  \vspace{-0.2cm}
\end{center}

In this paper we compute the QCD corrections for the triple Higgs boson production cross section via gluon fusion, within the heavy-top approximation.
We present, for the first time, analytical results for the next-to-leading order corrections, and also compute the soft and virtual contributions of the next-to-next-to-leading order cross section.
We provide predictions for the total cross section and the triple Higgs invariant mass distribution.
We find that the QCD corrections are large at both perturbative orders, and that the scale uncertainty is substantially reduced when the second order perturbative corrections are included.

\hfill

\end{titlepage}
\setcounter{footnote}{1}
\renewcommand{\thefootnote}{\fnsymbol{footnote}}

%
\section{Introduction}

After the discovery of a Higgs boson \cite{Aad:2012tfa,Chatrchyan:2012ufa} at the Large Hadron Collider (LHC), it is crucial to study its properties in order to determine whether it is indeed the particle predicted by the Standard Model (SM) or not.
Besides its couplings to fermions and gauge bosons, which are so far compatible with the SM expectations, it is of great interest to determine the Higgs boson self-couplings, which will allow to shed light on the scalar potential, and therefore the electroweak symmetry breaking mechanism.

The Higgs boson trilinear and quartic self-couplings $\lambda_3$ and $\lambda_4$ can be studied in hadron colliders via double and triple Higgs production, respectively (see Ref. \cite{Degrassi:2016wml} for an alternative method based on single Higgs production).
The SM expectations for these processes, corresponding to $\lambda_3=\lambda_4=m_H^2/(2v^2)$, being $v\simeq 246\text{ GeV}$ the Higgs vacuum expectation value and $m_H$ its mass, are very low.
For a collider energy of $14\text{ TeV}$, the leading order (LO) predictions for the double and triple Higgs production cross sections are of ${\cal O}(20\text{ fb})$ and ${\cal O}(0.05\text{ fb})$.
As a consequence, in a SM-like scenario, a determination of the trilinear coupling will be very challenging at the LHC, while the measurement of the quartic coupling via triple Higgs boson production will be at best relegated to a future collider \cite{Plehn:2005nk,Binoth:2006ym}.
Of course, the situation can be largely modified in the presence of new physics scenarios for the Higgs sector.

As it also happens for single and double Higgs production, the triple-Higgs final state is mainly produced in the SM via gluon fusion, mediated by a heavy quark (mainly top) loop.
For this production mechanism, the corresponding cross section is only known at LO in perturbation theory.
However, the QCD corrections are expected to be large, as it also happens for the other gluon initiated loop-induced processes mentioned above.
Unfortunately, their computation is very difficult. For instance, the next-to-leading order (NLO) corrections for double Higgs production (a simpler process with one particle less in the final state) were not known until very recently \cite{Borowka:2016ehy}.
In the absence of a full NLO calculation, and in order to provide an estimate of the size of the perturbative corrections, approximate NLO predictions were obtained in Ref. \cite{Maltoni:2014eza}, where only the exact real corrections were included.

In this paper, we present the first calculation of the QCD perturbative corrections for triple Higgs production within the Higgs effective field theory (HEFT). 
Within this framework, which formally corresponds to the large top quark mass limit of the SM, the Higgs bosons couple directly to gluons via an effective Lagrangian.
This approach has been successfully used to compute the QCD corrections for single and double Higgs production.
Motivated by this, we apply it to compute the NLO corrections and the next-to-next-to-leading order (NNLO) soft and virtual contributions for the total cross section and the triple Higgs system invariant mass distribution.

This work is organized as follows: in Section \ref{sec:virtuals} we present the virtual corrections up to NNLO,
and later in Section \ref{sec:reals}, after combining with the corresponding real corrections, we present the NLO and NNLO partonic cross sections, the latter within the soft-virtual approximation.
In Section \ref{sec:pheno} we present the numerical results for the LHC and future colliders, and compare our  predictions with the other NLO approximation available.
Finally, in Section \ref{sec:conc} we present our conclusions.

\section{Virtual corrections up to NNLO}
\label{sec:virtuals}

In this section we present the one and two-loop corrections to the triple-Higgs boson production cross section in hadronic collisions via gluon fusion.
As was stated before, we work within the HEFT were the Higgs bosons couple directly to gluons via the effective Lagrangian
\beq
{\cal L}_\text{eff} =
-\f{1}{4}G_{\mu\nu}^a G^{\mu\nu}_a
\left(
C_H \f{H}{v}
-C_{HH} \f{H^2}{2v^2}
+C_{HHH} \f{H^3}{3v^3}
+\dots
\right)\,,
\eeq
and where the matching coefficients can be expanded in powers of the strong coupling $\as$ as
\beq
C_X = -\f{\as}{3\pi}
\sum_{n\geq 0} C_X^{(n)} \left( \f{\as}{\pi} \right)^n\,.
\eeq
The three coefficients needed for our calculation are known up to fourth order in their perturbative expansion \cite{Chetyrkin:1997iv,Chetyrkin:2005ia,Kramer:1996iq,Schroder:2005hy,
Djouadi:1991tka,Grigo:2014jma,Spira:2016zna}.

For the generation of the relevant Feynman diagrams we employed {\sc qgraf} \cite{Nogueira:1991ex}, while the manipulation of the resulting amplitudes was performed with in-house routines written for {\sc Mathematica}.
Finally, we reduced the result into master integrals using the algorithm {\sc Fire} \cite{Smirnov:2008iw}.
The infrared divergent results were handled using dimensional regularization with $D=4-2\ep$ dimensions.

The virtual corrections to the partonic cross section can be written in terms of the squared matrix element as
\beq
\hat\sigma_v =
\f{1}{2 s}
\f{1}{3! 2^2 8^2 (1-\ep)^2}
\int |\overline{\cal M}|^2
\dPS
\equiv
\int d\hat\sigma_v\, \dPS
\eeq
where we include the flux factor, the average over initial state colors and helicities, and the factor $1/3!$ arising from the identical particles in the final state.
Here $\dPS$ represents the three particle phase space.
Expanding in powers of the strong coupling, we have
\beq
d\hat\sigma_v=
\left(\f{\as}{2\pi}\right)^2
\left[
d\hat\sigma^{(0)}+\f{\as}{2\pi}d\hat\sigma^{(1)}+\left(\f{\as}{2\pi}\right)^2 d\hat\sigma^{(2)}
+{\cal O}(\as^3)\right]\,.
\eeq

Exploiting the well known one and two-loop infrared behaviour of QCD amplitudes \cite{Catani:1998bh,Sterman:2002qn,Aybat:2006wq}, we can write the renormalized NLO and NNLO virtual corrections as
\beeq\label{eq:infrared}
d\hat\sigma^{(1)}\!\!\!&=&\!\!\!
2\,\text{Re}\left[\bom{I}_g^{(1)}\right]
d\hat\sigma^{(0)}
+d\hat\sigma^{(1)}_{\text{fin}}
\,,
\\[1.0ex]
\nn
d\hat\sigma^{(2)}\!\!\!&=&\!\!\!
\left(
\left|\bom{I}_g^{(1)}\right|^2 
+2\text{Re}\left[\left(\bom{I}_g^{(1)}\right)^2\right]
+2\text{Re}\left[\bom{I}_g^{(2)}\right]\right)
d\hat\sigma^{(0)}\!\!
+
2\,\text{Re}\left[\bom{I}_g^{(1)}\right]d\hat\sigma^{(1)}_{\text{fin}}
+d\hat\sigma^{(2)}_{\text{fin}}\,,
\eeeq
where $\bom{I}^{(1)}$ and $\bom{I}^{(2)}$ represent the one and two-loop insertion operators defined, for instance, in Ref.~\cite{Catani:1998bh}.

The $D$ dimensional LO cross section can be written as
\beq
d\hat\sigma^{(0)} =
F_{LO}^D |C_{LO}^{3H}|^2 (1-\ep),
\;\;\;\;\;
\text{with}
\;\;\;\;\;
F_{LO}^D = 
\f{s}{1728 v^6 (1-\ep)^2}\,,
\eeq
and where the coefficient $C_{LO}^{3H}$ is defined as \beeq\label{eq:C3HLO}
C_{LO}^{3H} &=&
2
+\frac{6 \lambda _4 v^2}{s_{345}-m_H^2}
 +\left( \f{36\lambda _3^2 v^4}{s_{345}-m_H^2}-6\lambda _3 v^2\right)
   \left(\frac{1}{s_{35}-m_H^2}+\frac{1}{s_{45}-m_H^2}+\frac{1}{s_{34}-m_H^2}\right) .
\eeeq
Here $s_{ij\dots k}=(p_i + p_j + \dots +p_k)^2$.
For simplicity, we have set $\Gamma_H = 0$ (the numerical effect due to the  Higgs width is negligible), in which case the $C_{LO}^{3H}$ coefficient is a real number.

\begin{figure}
\begin{center}
\begin{picture}(300,190)
\put(0,0){
\includegraphics[width=10cm]{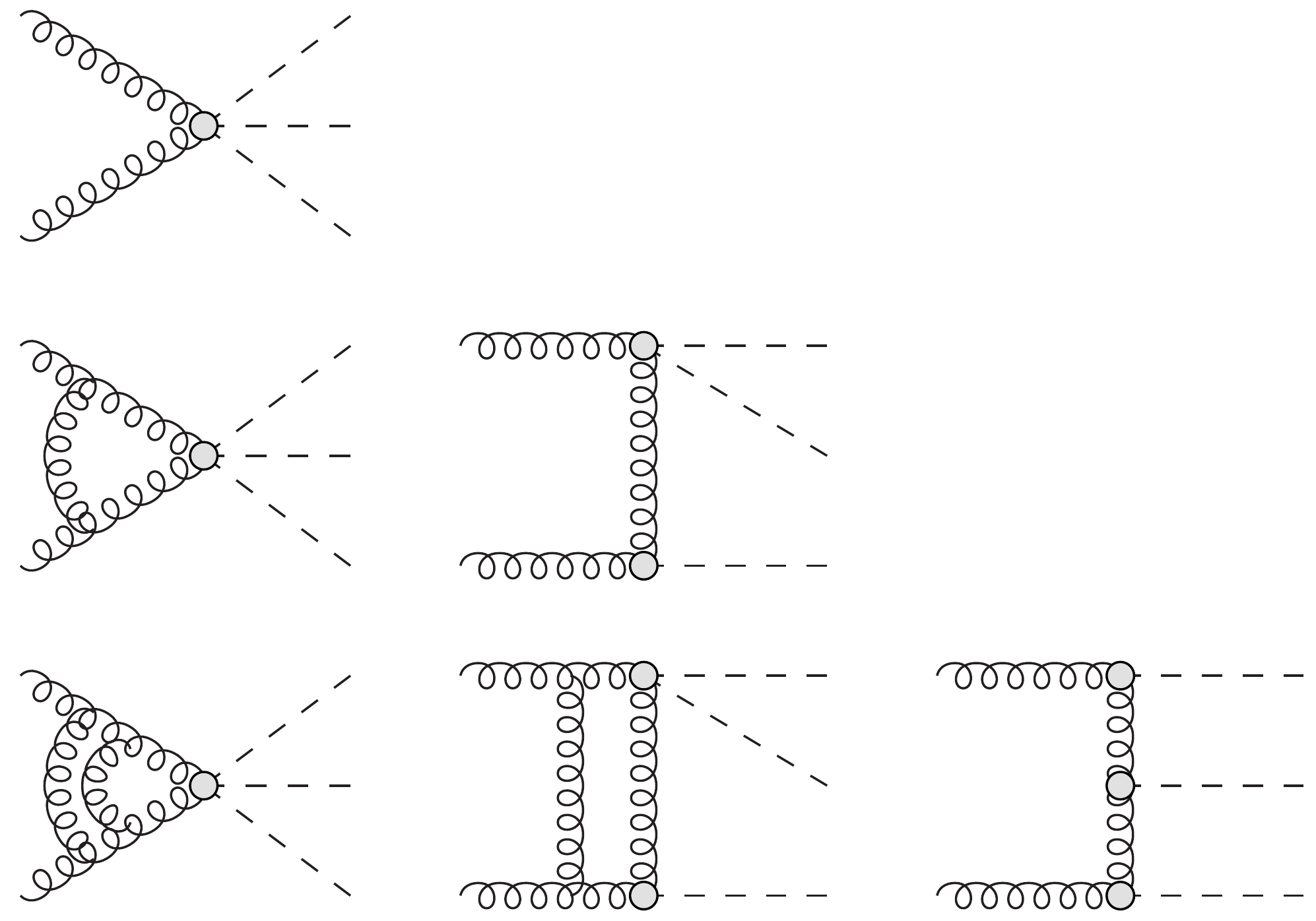}
}
\put(-30,167){${\cal O}(\as)$}
\put(-30,96){${\cal O}(\as^2)$}
\put(-30,25){${\cal O}(\as^3)$}
\end{picture}
\end{center}
\vspace*{-0.5cm}
\caption{\label{fig:diagrams}\small
Examples of the different Feynman diagrams contributing to the triple Higgs production virtual corrections at the different perturbative orders.
}
\end{figure}
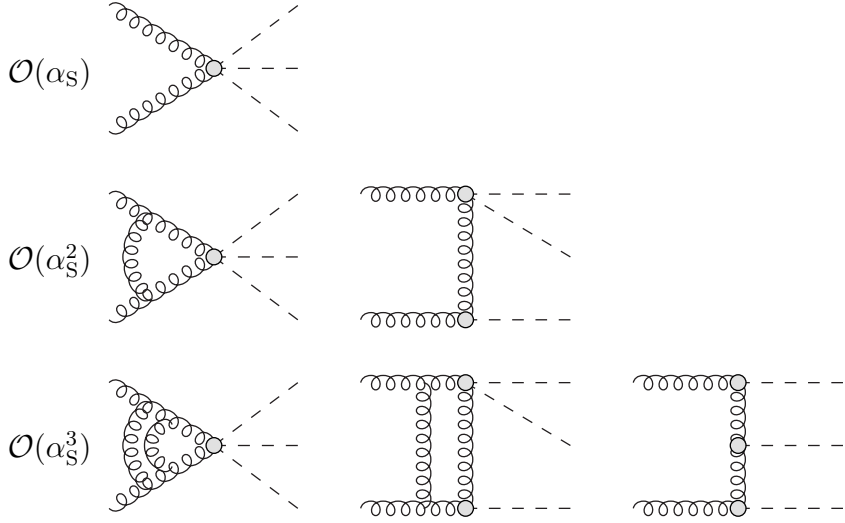

The one and two-loop infrared-regulated parts can be organized in the following way:
\beeq\label{eq:finite}
d\hat\sigma^{(1)}_{\text{fin}}&=&F_{LO}^D
\left\{
|C_{LO}^{3H}|^2\,{\cal F}^{(1)}
+\text{Re}(C_{LO}^{3H})\,{\cal R}^{(1)}_{3H}
+{\cal O}(\ep^3)
\right\}\,,\\[1.0ex]
d\hat\sigma^{(2)}_{\text{fin}}&=&F_{LO}^D
\left\{
|C_{LO}^{3H}|^2 \,{\cal F}^{(2)}
+\text{Re}(C_{LO}^{3H}) 
\left(
{\cal R}^{(2)}_{3H}
+ \,{\cal S}^{(2)}_{3H}
+ \,{\cal T}^{(2)}_{3H}
\right)
+ \,{\cal V}^{(2)}_{3H}
+{\cal O}(\ep)
\right\}\,.\nn
\eeeq
The contributions labelled ${\cal F}$ arise from diagrams with only one HEFT operator insertion.
The ones in ${\cal R}$ originate from the interference between amplitudes with two HEFT operator insertions and amplitudes with only one insertion.
On the other hand, contributions in ${\cal T}$ arise from the interference between diagrams with three operator insertions and the LO, and the ones in ${\cal V}$ come from the square of amplitudes with two insertions.
Finally, contributions to ${\cal S}$ have their origin on the difference between the NNLO QCD corrections to the effective vertices $Hgg$, $HHgg$ and $HHHgg$.
In Figure \ref{fig:diagrams} we show illustrative examples of the different Feynman diagrams involved in the calculation of the virtual corrections.
As already mentioned, since we adopted the Higgs zero-width approximation, both $C_{LO}^{3H}$ and $C_{LO}^{2H}$ are real numbers.
Beyond that limit, there is also a numerically negligible contribution proportional to $\text{Im}(C_{LO}^{2H}) \text{Im}(C_{LO}^{3H})$, which we will ignore.

We start by presenting the one-loop corrections.
For simplicity, we set $\mu_R = \mu_F = Q$ throughout the rest of the work, being $Q$ the invariant mass of the triple-Higgs system.
We find that
\beeq
{\cal F}^{(1)}&=&
11
+\ep
   \left(\frac{7}{6} \zeta_2
   (2 N_f-33)+12 \zeta_3-17\right)\nn\\
   &+&\ep^2 \bigg(
\frac{7}{6} \zeta_2 (33-2 N_f)+   
   \frac{1}{9} \zeta_3
   (2 N_f-141)
   +18\zeta_4-12\bigg)\,,\nn
   \\[2ex] 
   {\cal R}^{(1)}_{3H} &=&
C_{LO}^{2H}(s_{34}) \,
r^{(1)}(s,t_{15},t_{25},m_H^2,s_{34}) 
+
C_{LO}^{2H}(s_{35}) \,
r^{(1)}(s,t_{14},t_{24},m_H^2,s_{35})\nn\\[1.5ex]
&+&
C_{LO}^{2H}(s_{45}) \,
r^{(1)}(s,t_{13},t_{23},m_H^2,s_{45}) \,,\label{eq:R1}
\eeeq
where $N_f$ represents the number of light partons, $\zeta_n$ stands for the Riemann zeta function, and
\beq\label{eq:r1small}
r^{(1)}(s,t,u,m_1^2,m_2^2) =
\f{4}{3} + \f{2\ep}{3 s}\left[
m_1^2 m_2^2 \left(\f{1}{t} + \f{1}{u} \right)
- (s + m_1^2 + m_2^2)
\right] 
\eeq
with $t_{ij} = (p_i - p_j)^2$, and where we have defined
\beq
C_{LO}^{2H}(s) =
\f{6 v^2 \lambda_3}{s - m_H^2} - 1\,.
\eeq
Notice that, given that these contributions enter in the two-loop result multiplied by a double pole, we need their expansion up to ${\cal O}(\ep^2)$.

We start now with the NNLO results.
For the coefficient ${\cal F}$ we find
\beeq
{\cal F}^{(2)}&=&
\left(\frac{8 N_f}{3}+\frac{19}{2}\right) \log\left(\f{s}{m_t^2}\right)
+N_f \left(\frac{217
   \zeta_2}{12}-\frac{17
   \zeta_3}{6}-\frac{3239}{108}\right)\\[1.0ex]
&-&\frac{11  \zeta_2 N_f^2 }{18}
   -\frac{249
   \zeta_2}{2}-\frac{253 \zeta_3}{4}+\frac{45
   \zeta_4}{8}+\frac{8971}{36}\,,\nn
   \eeeq
where $m_t$ stands for the top quark mass.
The function ${\cal R}_{3H}$ can be written at this order as
\beeq
{\cal R}^{(2)}_{3H} &=&
C_{LO}^{2H}(s_{34}) \,
r^{(2)}(s,t_{15},t_{25},m_H^2,s_{34}) 
+
C_{LO}^{2H}(s_{35}) \,
r^{(2)}(s,t_{14},t_{24},m_H^2,s_{35})\nn\\ [1.5ex]
&+&
C_{LO}^{2H}(s_{45}) \,
r^{(2)}(s,t_{13},t_{23},m_H^2,s_{45}) \,,
\eeeq
where
{\small
\beeq\label{eq:r2small}
r^{(2)}(s,t,u,m_1^2,m_2^2) &=&
-2\left(
1 + \f{m_1^4 + m_2^4}{s^2}
\right)
\Bigg\{
 \log \left(-\frac{m_1^2}{t}\right) \log
   \left(1-\frac{m_1^2}{t}\right)+\log \left(-\frac{m_2^2}{t}\right) \log
   \left(1-\frac{m_2^2}{t}\right)
\nn\\&&\hspace*{-3cm}   
   +\log \left(-\frac{m_1^2}{u}\right) \log
   \left(1-\frac{m_1^2}{u}\right)+\log \left(-\frac{m_2^2}{u}\right) \log
   \left(1-\frac{m_2^2}{u}\right)-\f{1}{2}\log ^2\left(\frac{u}{t}\right)
   \nn\\&&\hspace*{-3cm}
   +  
   \text{Li}_2\left(1-\frac{m_1^2 m_2^2}{t
   u}\right)+\text{Li}_2\left(\frac{m_1^2}{t}\right)+\text{Li}_2\left(\frac{m_2^
   2}{t}\right)+\text{Li}_2\left(\frac{m_1^2}{u}\right)+\text{Li}_2\left(\frac{m
   _2^2}{u}\right)
   \Bigg\} 
   \nn\\&&\hspace*{-3cm}
+ \f{4 \pi ^2}{3 s^2}
   \left(m_1^4+m_2^4\right)
    - \f{ 1 }{9 } \left(33 - 2 N_f\right)  \log \left(\frac{t u}{s^2}\right)
+  \f{2 }{27 } \left(-10 N_f+18
   \pi ^2+471\right)
   +\f{2}{s} \left(m_1^2+m_2^2\right)  \,.
\eeeq
}
For ${\cal S}_{3H}^{(2)}$ we have
\beq
{\cal S}^{(2)}_{3H} =
16
\left[
(C_{3H}^{(2)} - C_{H}^{(2)})
 - 3 \lambda_3 v^2
 (C_{2H}^{(2)} - C_{H}^{(2)})
 \left(
 \f{1}{s_{34}-m_H^2}
+ \f{1}{s_{35}-m_H^2}
+ \f{1}{s_{45}-m_H^2} 
 \right)
\right]\,,
\eeq
where the NNLO corrections to the effective vertices between Higgs bosons and gluons imply
\cite{Chetyrkin:1997iv,Chetyrkin:2005ia,Kramer:1996iq,Schroder:2005hy,
Djouadi:1991tka,Grigo:2014jma,Spira:2016zna}
\beq
C_{2H}^{(2)} - C_{H}^{(2)}
=
\tfrac{2}{3} (C_{3H}^{(2)} - C_{H}^{(2)})
=
35/24 + 2 N_f/3 \,.
\eeq
For the function ${\cal T}^{(2)}_{3H}$ we write
\beq
{\cal T}^{(2)}_{3H} = H(3,4) + H(3,5) + H(4,5) \,,
\eeq
where we have defined
\beeq
H(A,B) &=&
\f{1}{9 s^2 t_{1A} t_{2 B}}
\bigg\{
s^2
   \left(s_{A B}^2+2 t_{1A} t_{2 B}\right)
 -2 s t_{2 A} t_{1B} s_{A B}+\left(t_{2 A} t_{1B}-t_{1A} t_{2 B}\right){}^2+8 s m_H^6
\nn\\[1.0ex]
&+&m_H^4 \left(-2 s \left(s_{A B}+3 \left(t_{2 A}+t_{1B}\right)+t_{1A}+t_{2 B}\right)+\left(t_{1A}-t_{2 A}-t_{1B}+t_{2 B}\right){}^2+2 s^2\right) 
\nn\\[1.5ex]
&+&2 m_H^2 \big(-2 s^2 s_{A B}+s
   \left(s_{A B} \left(t_{2 A}+t_{1B}\right)+2 t_{2 A} t_{1B}+t_{1A} t_{2 A}+t_{1B} t_{2 B}\right)
\nn\\[1.0ex]
   &-&\left(t_{1A}-t_{2 A}-t_{1B}+t_{2 B}\right) \left(t_{1A} t_{2 B}-t_{2 A} t_{1B}\right)\big)  
\bigg\} + (1\leftrightarrow 2) \,,
\eeeq
where the last term indicates that the momenta $p_1$ and $p_2$ have to be exchanged.
Finally, the function ${\cal V}_{3H}^{(2)}$ can be expressed as
\beeq 
{\cal V}^{(2)}_{3H} &=&
F(3+4,5,3+4,5)+F(3+5,4,3+5,4)+F(4+5,3,4+5,3) \nn\\[1.0ex]
&+& 2\left[F(3+4,5,3+5,4)+F(3+4,5,4+5,3)+F(3+5,4,4+5,3)\right] \,,
\eeeq
with the following definitions,
\beq
F(A,B,C,D) =
C_{LO}^{2H}(m_A^2)
C_{LO}^{2H}(m_C^2)
[
G(A,C) + G(A,D) + G(B,C) + G(B,D)
]
\eeq
and
\beeq
&& G(A,C)=\f{1}{18 s^2 t_{1A} t_{1C}}
\bigg\{
\left(s \left(t_{1A}+s_{AC}\right)+m_A^2 \left(-m_C^2+t_{1C}-2
   s\right)+m_C^2 \left(t_{2 A}-s\right)-t_{2 A} t_{1C}\right){}^2\nn\\[1.0ex]
   &&\;\;\;+2 s
   \left(m_A^2-t_{1A}\right) \left(\left(m_C^2-t_{2 C}-s\right) \left(2
   m_A^2-t_{1A}-s_{AC}+2 m_C^2-t_{1C}\right)-t_{1C} \left(-m_A^2+t_{2
   A}+s\right)\right)\nn\\[1.0ex]
   &&\;\;\;+\left(m_A^2-t_{1A}\right){}^2 \left(-m_C^2+t_{2
   C}+s\right){}^2-2 s t_{1A} \left(m_C^2-t_{1C}\right) \left(-m_C^2+t_{2
   C}+s\right)
\bigg\} \,.
\eeeq
Here the Lorentz invariants are defined as
\beq
t_{iX} = (p_i - p_X)^2\,, \;\;\;\;\;
s_{XY} = (p_X + p_Y)^2\,, \;\;\;\;\;
m_X^2  =  p_X^2\,,
\eeq
with $p_{i+j} = p_i + p_j$.

With the above results we complete the presentation of the NNLO virtual corrections for the triple Higgs boson production cross section.
It is worth to mention that some of the expressions obtained can be directly related to their analogous in double Higgs production.
This is of course the case of the ${\cal F}$ contributions, which arise directly from the gluon form factor, and take exactly the same value for both processes.
Less trivially, the functions $r^{(1)}$ and $r^{(2)}$ defined in Eqs. (\ref{eq:r1small}) and (\ref{eq:r2small}) are equal to the coefficients ${\cal R}^{(1)}$ and ${\cal R}^{(2)}$ defined in Eqs.~(11) and (15) of Ref.~\cite{deFlorian:2013uza}, provided that the limit $m_1 = m_2 = m_H$ is taken in the former.

\section{NLO and NNLO$_\text{SV}$ partonic cross sections}
\label{sec:reals}

We present here the partonic cross section $\hat\sigma$ at NLO and NNLO, obtained by combining the results from the previous section with the corresponding real corrections (computed within the soft approximation for the NNLO case).
We recall that the inclusive hadronic cross section can be obtained from the partonic result as
\beq
\f{d\sigma}{dQ^2} =
\sum_{i,j}
\int_0^1 dx_1 dx_2 f_{i/h_1}(x_1)f_{j/h_2}(x_2)
\int_0^1 dx \,\delta\left(x-\f{Q^2}{x_1 x_2 s_H}\right)
\f{d\hat\sigma_{ij}}{dQ^2}\,,
\eeq
where $\sqrt{s_H}$ represents the collider center-of-mass energy.
The parton densities are denoted by $f_{i/h}(x)$ and the subscripts $i,\, j$ label the type of massless partons.
For simplicity, the dependence on the factorization and renormalization scales is always understood.

The partonic cross section $\hat\sigma$ can be expanded in powers of the strong coupling $\as$.
Up to NNLO, we have
\beq
Q^2 \f{d\hat\sigma_{ij}}{dQ^2} = \hat\sigma_{LO}
\left\{
\eta_{ij}^{(0)}
+ \left(\f{\as}{2\pi}\right) \eta_{ij}^{(1)}
+ \left(\f{\as}{2\pi}\right)^2 \eta_{ij}^{(2)}
+ {\cal O}(\as^3)
\right\}\,,
\eeq
where the LO cross section in the HEFT takes the form
\beq
\hat\sigma_{LO} = \int \dPS \left(\f{\as}{2\pi}\right)^2 F_{LO} \left|C_{LO}^{3H}\right|^2 
\;\;
\text{ with }
\;\;
F_{LO} = 
\f{Q^2}{1728 v^6}\,,
\eeq
and where $C_{LO}^{3H}$ is defined in Eq.~(\ref{eq:C3HLO}).

At LO only the gluon initiated subprocess contributes, and therefore we have
\beq
\eta^{(0)}_{ij} = \delta(1-x)\delta_{ig}\delta_{jg}\,,
\eeq
where $x=Q^2/s$, being $\sqrt{s}$ the partonic center-of-mass energy.

For the calculation of the NLO results, we exploited the relation that can be established between some contributions to the triple-Higgs boson cross section and the single-Higgs boson one, as was already discussed for the double-Higgs case in Ref.~\cite{deFlorian:2013jea}.
We find the following results,
\beeq
\eta^{(1)}_{gg} &=& 
\left ( 11 + 12 \zeta_2 +  
{\cal R}^{(1)}_{3H}\, \f{\text{Re}\left(C_{LO}^{3H}\right)}
{ \left|C_{LO}^{3H}\right|^2}
  \right ) \delta(1-x) 
+ 24  {\cal D}_1(x)
\nonumber \\
&-& 24x(-x+x^2+2)\ln(1-x)
-\frac{12 (x^2+1-x)^2}{1-x}\ln(x)
-11 (1-x)^3, \label{eq:eta_gg1}
\\
\eta^{(1)}_{qg} &=& 
-\frac{4}{3} \left ( 1+(1-x)^2 \right )\ln \frac{x}{(1-x)^2}-2+4x
-\frac{2}{3}x^2, \\
\eta^{(1)}_{q \bar q}  &=& \frac{64}{27}(1-x)^3,
\eeeq
where the plus distributions ${\cal D}_i(x)$ are defined as usual,
\beq
{\cal D}_i(x) = \left[
\f{\ln^i (1-x)}{(1-x)}
\right]_+\,,
\eeq
and the coefficient ${\cal R}^{(1)}_{3H}$ is defined in Eq.~(\ref{eq:R1}).
Since we are dealing in this section with finite quantities, the $\ep=0$ limit can be taken for this coefficient.

The results above complete the NLO calculation within the HEFT which, to the best of our knowledge, has not been presented before.
For the NNLO$_\text{SV}$ cross section, we make use of the universal formula derived in Ref. \cite{deFlorian:2012za}.
The soft and virtual contributions are those proportional to the delta function $\delta(1-x)$ and the plus distributions ${\cal D}_i(x)$, which in Mellin space correspond to constants and threshold enhanced logarithms.
These contributions only appear in the gluon initiated partonic channel, and they can be expressed as
\beeq
\eta_{gg\text{(SV)}}^{(2)} &=&
\delta(1-x)
\Bigg[
\frac{11}{18} \zeta _2 N_f^2+\left(-\frac{99 \zeta _2}{4}+\frac{37
   \zeta _3}{6}-\frac{82}{27}\right) N_f+\frac{517 \zeta
   _2}{2}-\frac{407 \zeta _3}{4}-\frac{81 \zeta _4}{8}+\frac{607}{9} \nn\\
   &+& 12 \zeta _2
   \f{d\hat\sigma_\text{fin}^{(1)}}{d\hat\sigma^{(0)}}+\f{d\hat\sigma_\text{fin}^{(2)}}{d\hat\sigma^{(0)}}
\Bigg]
+ {\cal D}_0 (x) \left[
\left(\frac{56}{9}-8 \zeta _2\right) N_f+132 \zeta _2+702 \zeta
   _3-\frac{404}{3}
\right]\nn\\
&+& {\cal D}_1 (x) \left[
-\frac{40 N_f}{3}-360 \zeta _2+268+24 \f{d\hat\sigma_\text{fin}^{(1)}}{d\hat\sigma^{(0)}}
\right]
+ {\cal D}_2 (x) \left(
8 N_f-132
\right)
+288 {\cal D}_3 (x)\,,
\eeeq
where the finite reminders of the one and two-loop virtual corrections $d\hat\sigma_\text{fin}$ are defined in Eq.~(\ref{eq:finite}).

As it was already observed in Refs. \cite{Catani:2003zt,deFlorian:2012za}, the SV approximation yields better results when defined in Mellin space.
To this end, we need to compute the following $N$-moments,
\beeq
f_{i/h,N} &=& \int_0^1 dx\, x^{N-1}\, f_{i/h}(x)\,, \\
\eta_{gg\text{(SV)},N}^{(2)}
&=&
\int_0^1 dx\, x^{N-1}\,
\eta_{gg\text{(SV)}}^{(2)}(x)\,, \\
\tilde\eta_{gg\text{(SV)},N}^{(2)} &=&
\eta_{gg\text{(SV)},N}^{(2)} \Big\vert_{\ln^k N,\text{ const.}}\,,
\eeeq
where we additionally take the large-$N$ limit on the resulting expression for the Mellin transform $\eta_{gg\text{(SV)},N}^{(2)}$, retaining only the logarithmically enhanced and constant terms.
Its explicit expression can be easily derived from the results in Ref. \cite{deFlorian:2012za}.
Finally, the NNLO contribution to the physical cross section in the SV approximation can be obtained by Mellin inversion,
\beq
Q^2 \f{d\sigma_{gg}^{(2)}}{dQ^2} =
\hat\sigma_{LO}
\left(\f{\as}{2\pi}\right)^2
\int_{C_{MP}-i\infty}^{C_{MP}+i\infty}
\f{dN}{2\pi i} \left(\f{Q^2}{s_H}\right)^{-N+1}
f_{g/h_1,N} \, f_{g/h_2,N} \,
\tilde\eta_{gg\text{(SV)},N}^{(2)}\,,
\eeq
where the constant $C_{MP}$ defining the contour of integration is on the right of all the singularities of the integrand, as defined in the Minimal Prescription in Ref. \cite{Catani:1996yz}.

\section{Phenomenological results}
\label{sec:pheno}

We present in this section our numerical predictions for the LHC and future hadron colliders.
The NNLO corrections are computed within the soft-virtual (SV) approximation, which has proven to be an excellent estimation to the full HEFT result for other gluon-initiated processes, like single and double Higgs boson production \cite{Catani:2001ic,deFlorian:2013uza}.
In particular, for the triple Higgs production cross section, we find that at NLO the SV approximation differs from the full HEFT result by less than $2.5\%$.

In order to partially retain the dependence on the top quark mass, we normalize the QCD corrections computed in the HEFT with the exact LO result, differentially in the triple Higgs system invariant mass.
The full (loop-induced) LO cross section was computed using {\sc MadGraph5\_aMC@NLO} \cite{Alwall:2014hca,Hirschi:2015iia}.

For the numerical implementation we set the values $m_H = 125\text{ GeV}$ and $m_t = 173.2\text{ GeV}$ for the Higgs boson and top quark masses.
We use the MMHT2014 sets \cite{Harland-Lang:2014zoa} for the parton flux and strong coupling, at the corresponding order for the LO, NLO and NNLO predictions.
For the renormalization and factorization scales we use two different central scale choices, $\mu_R = \mu_F = \mu_0$ with $\mu_0 = Q$ and $\mu_0 = Q/2$.
As usual, the theoretical uncertainty arising from the missing higher orders is estimated by varying these scales independently in the range $[\mu_0/2;2\mu_0]$, with the constraint $0.5<\mu_R/\mu_F<2$.

\begin{figure}
\begin{center}
\begin{tabular}{c c}
\begin{tabular}{c}
\includegraphics[width=8cm]{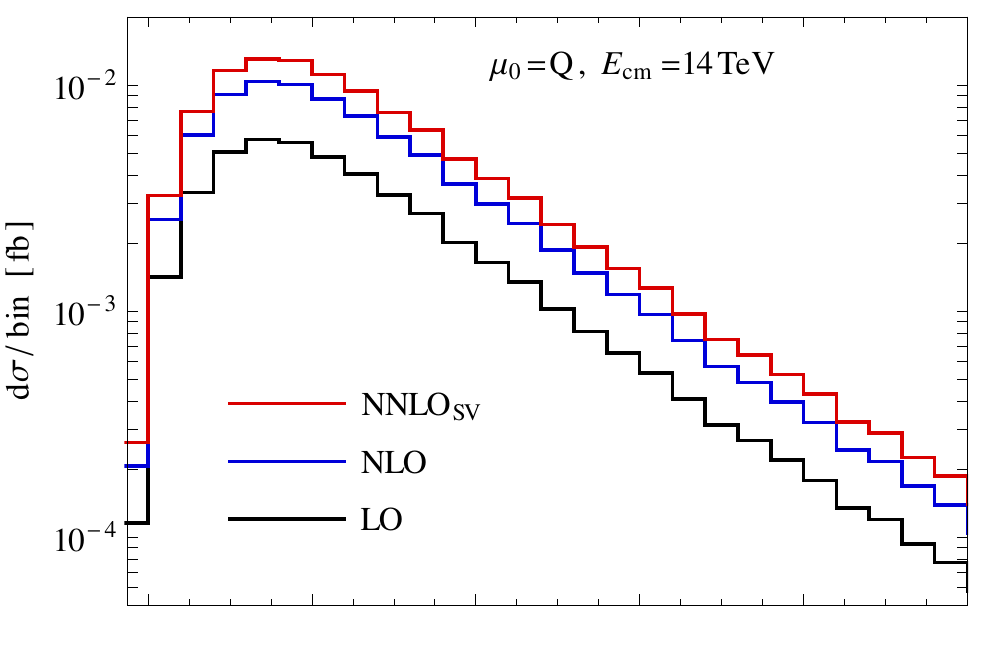}
\vspace*{-0.45cm}
\\
\hspace*{-0.08cm}
\includegraphics[width=7.79cm]{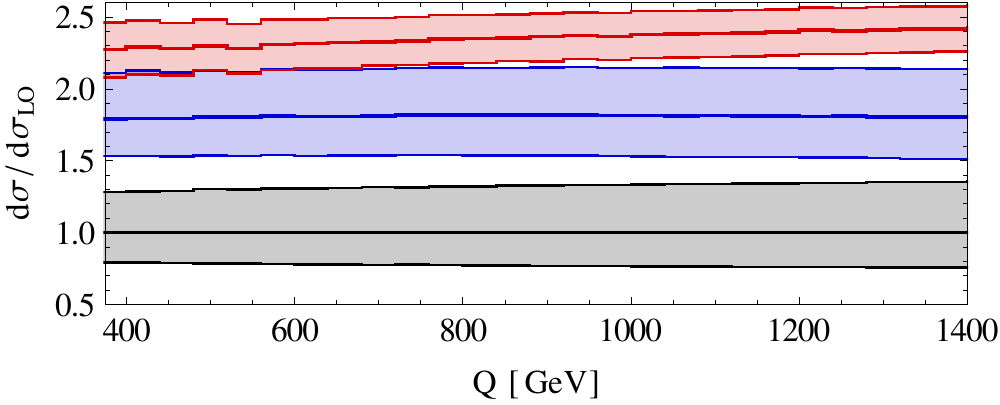}
\end{tabular}
&
\begin{tabular}{c}
\includegraphics[width=8cm]{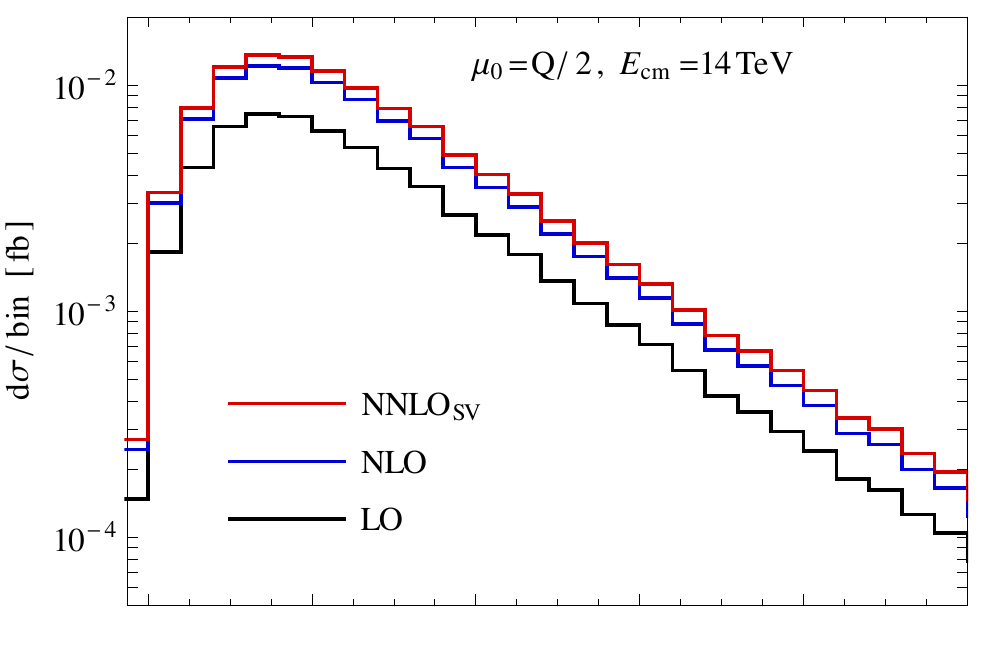}
\vspace*{-0.44cm}
\\
\hspace*{-0.08cm}
\includegraphics[width=7.79cm]{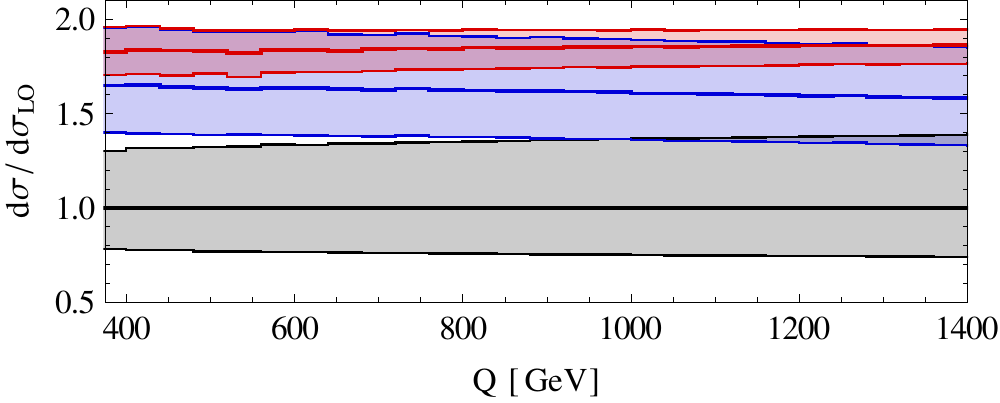}
\end{tabular}
\end{tabular}
\end{center}
\caption{\label{fig:14TeV_Q}\small
Triple Higgs invariant mass distribution for $E_{cm}=14\text{ TeV}$ at LO (black), NLO (blue) and NNLO$_\text{SV}$ (red), for the central scales $Q$ (left) and $Q/2$ (right).
The lower panel shows the ratio with respect to the LO, together with the scale uncertainty.
}
\end{figure}

In Figure \ref{fig:14TeV_Q} we show the triple Higgs system invariant mass distribution for a collider energy of $14\text{ TeV}$, for the two central scale choices.
As can be seen in the lower panel, for both choices the NLO corrections turn out to be large, with almost flat $K$-factors which approximately take the values $1.8$ and $1.6$ for $\mu_0 = Q$ and $\mu_0 = Q/2$, respectively.
The relative scale uncertainty is reduced at NLO, but still remains rather large,  $\sim 33\%$ for both scales. 
It is worth to notice that the scale variation at LO fails to anticipate the size of the higher order corrections, as it occurs for single and double Higgs production as well.

The NNLO corrections are still very sizeable, specially for $\mu_0 = Q$, where they represent an increase of about $28\%$ with respect to the NLO.
Corrections are more moderate for $\mu_0 = Q/2$, increasing the NLO result by about $13\%$, and with a large overlap between the corresponding uncertainty bands.
The $K$-factors have a mild dependence on the invariant mass of the system, showing a small increase towards larger values of $Q$.
For both scale choices, the total scale uncertainty is substantially reduced at NNLO, being of about $15\%$ and $12\%$ for $\mu_0 = Q$ and $\mu_0 = Q/2$, respectively.
It is worth pointing out that at NNLO both scale choices give fully compatible results, with a difference between the central values smaller than $4\%$.

\begin{figure}
\begin{center}
\begin{tabular}{c c}
\begin{tabular}{c}
\includegraphics[width=8cm]{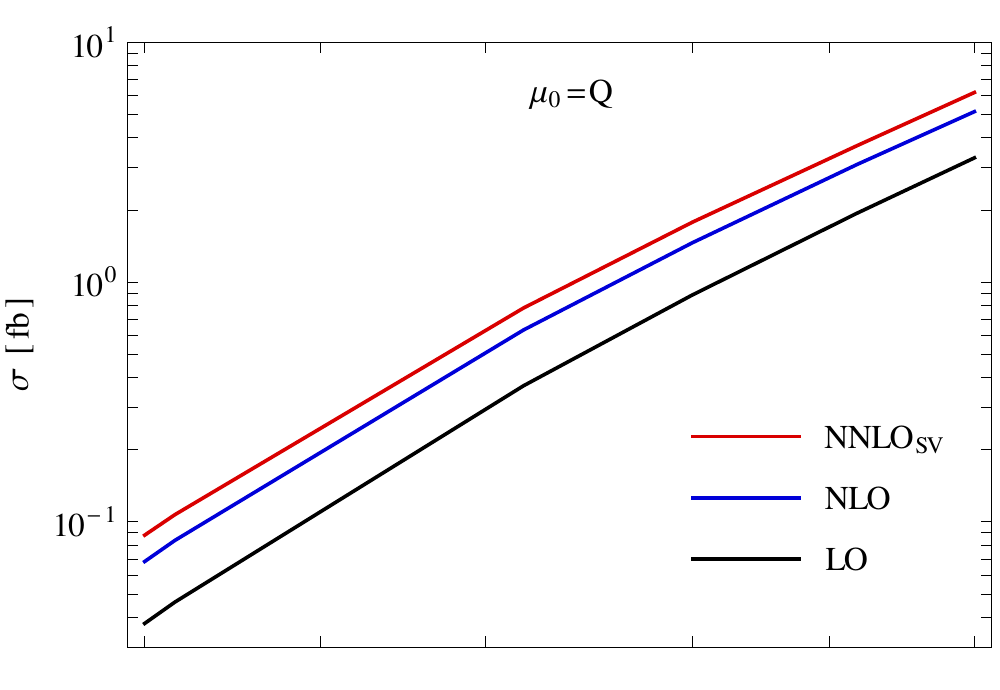}
\vspace*{-0.74cm}
\\
\hspace*{-0.09cm}
\includegraphics[width=7.79cm]{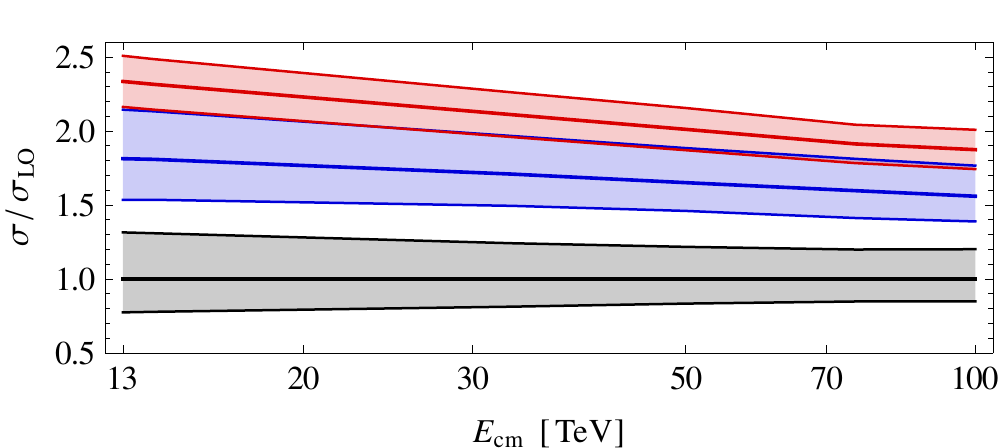}
\end{tabular}
&
\begin{tabular}{c}
\includegraphics[width=8cm]{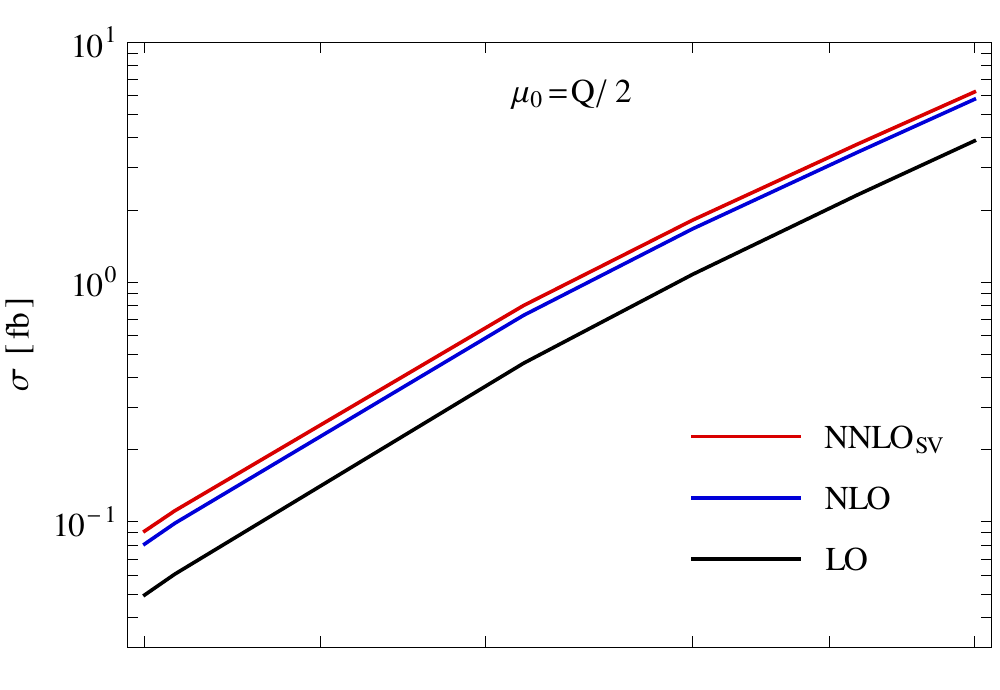}
\vspace*{-0.74cm}
\\
\hspace*{-0.09cm}
\includegraphics[width=7.79cm]{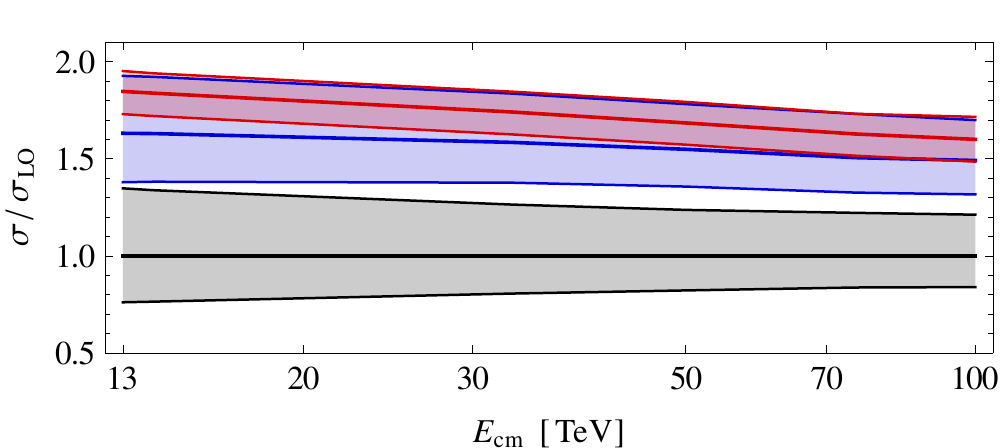}
\end{tabular}
\end{tabular}
\end{center}
\caption{\label{fig:totalXS}\small
Total cross section for the triple Higgs production as a function of the collider energy at LO (black), NLO (blue) and NNLO$_\text{SV}$ (red), for the central scales $Q$ (left) and $Q/2$ (right).
The lower panel shows the ratio with respect to the LO, together with the scale uncertainty.
}
\end{figure}

In Figure \ref{fig:totalXS} we show the triple Higgs production cross section at the different accuracy levels as a function of the collider energy.
Also, we present in Table \ref{table} the total cross section for $E_{cm} = 13,\,14,$ and $100\text{ TeV}$.
We can observe that both NLO and NNLO corrections are very sizeable in the whole range under study, taking lower values for higher energies.
Once again, the overlap between the NLO and NNLO predictions is larger for $\mu_0 = Q/2$.

\begin{table}
\begin{center}
\begin{tabular}{l | c c c}
\hline\hline
$\mu_0 = Q$ & $13\text{ TeV}$ & $14\text{ TeV}$ & $100\text{ TeV}$
\\
\hline
LO & $ 0.0377^{+31\%}_{-23\%} $ & $ 0.0462^{+31\%}_{-22\%} $ & $ 3.29^{+20\%}_{-15\%} $
\\[0.6ex]
NLO & $ 0.0683^{+18\%}_{-15\%} $ & $ 0.0836^{+18\%}_{-15\%} $ & $ 5.13^{+13\%}_{-11\%} $
\\[0.6ex]
NNLO$_{\text{SV}}$ & $ 0.0880^{+7.4\%}_{-7.4\%} $ & $ 0.107^{+7.4\%}_{-7.4\%} $ & $ 6.17^{+7.2\%}_{-7.0\%} $
\\
\hline\hline
$\mu_0 = Q/2$ & $13\text{ TeV}$ & $14\text{ TeV}$ & $100\text{ TeV}$
\\
\hline
LO & $ 0.0495^{+35\%}_{-24\%} $ & $ 0.0605^{+34\%}_{-24\%} $ & $ 3.88^{+21\%}_{-16\%} $
\\[0.6ex]
NLO & $ 0.0808^{+18\%}_{-15\%} $ & $ 0.0986^{+18\%}_{-15\%} $ & $ 5.78^{+14\%}_{-12\%} $
\\[0.6ex]
NNLO$_{\text{SV}}$ & $ 0.0914^{+5.7\%}_{-6.3\%} $ & $ 0.111^{+5.6\%}_{-6.4\%} $ & $ 6.20^{+7.2\%}_{-7.2\%} $
\\
\hline\hline
\end{tabular}
\end{center}
\caption{\label{table}\small
Triple Higgs boson production cross section (in $\text{fb}$) for different collider energies and at the different accuracy levels.
The uncertainties correspond to the scale variation.
}
\end{table}


From the results in Figure \ref{fig:totalXS}, we can also observe that the scale uncertainty is substantially reduced once the NNLO corrections are included, independently from the collider energy under consideration.
Specifically, at $13\text{ TeV}$ and for $\mu_0 = Q/2$ the total uncertainty goes from $59\%$ to $33\%$ and $12\%$ when going from LO to NLO and NNLO.
The analogue uncertainties at $100\text{ TeV}$ are $37\%$, $26\%$ and $14\%$, where we can also observe this reduction.
Similar results are obtained with $\mu_0 = Q$.

\begin{figure}
\begin{center}
\begin{tabular}{c c}
\includegraphics[width=8cm]{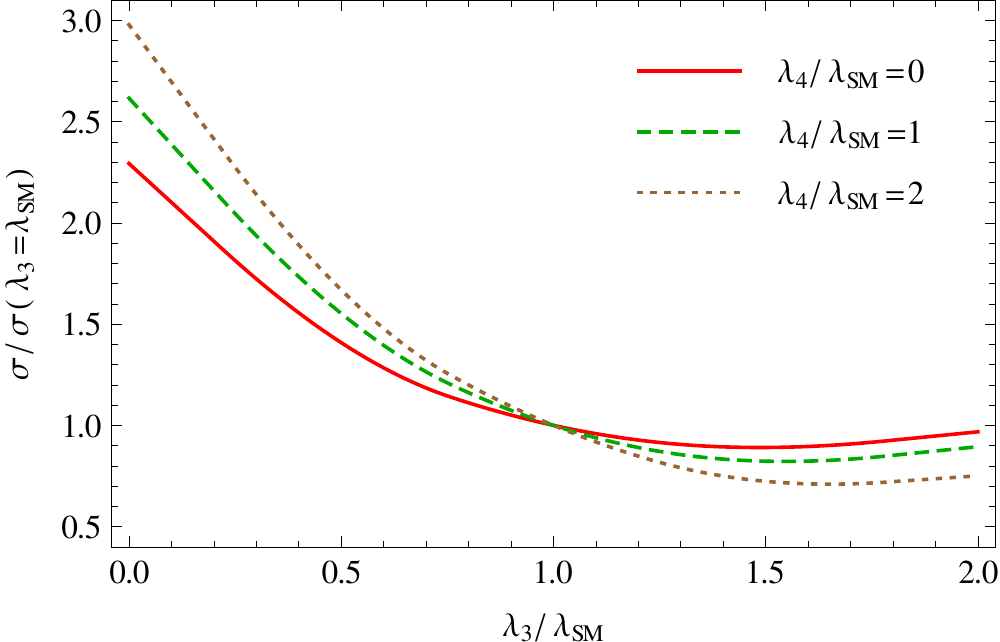}
&
\includegraphics[width=8cm]{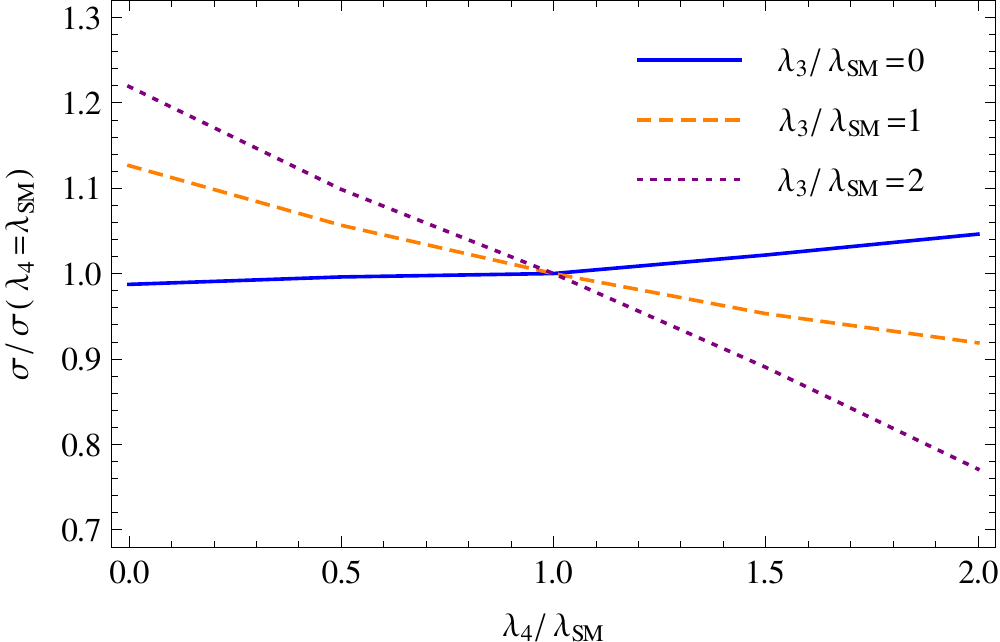}
\end{tabular}
\end{center}
\vspace*{-0.4cm}
\caption{\label{fig:lambda}\small
Dependence of the total NNLO cross section on the Higgs self-couplings $\lambda_3$ (left) and $\lambda_4$ (right), in units of the $\lambda_i = \lambda_\text{SM}$ prediction. 
}
\end{figure}

For completeness, we show in Figure \ref{fig:lambda} the dependence of the NNLO total cross section on the value of the Higgs self-couplings.
We do not intend to perform a full analysis for non-SM scenarios, but just to illustrate the sensitivity of this observable to $\lambda_3$ and $\lambda_4$ arround their SM expectations (in particular, in the range $\lambda/\lambda_\text{SM} \in [0;2]$), which is of course relevant for a future measurement. The results correspond to a center of mass energy of $100\text{ TeV}$, and the scale choice $\mu_R = \mu_F = Q/2$.
We can observe that, unfortunately, the dependence on $\lambda_4$ is rather small, and that the situation slightly improves for $\lambda_3>\lambda_{\text{SM}}$, while the sensitivity is even lower for $\lambda_3<\lambda_{\text{SM}}$.
It is worth to mention that the dependence of the ratio between the NNLO and LO cross sections on the value of the self-couplings is quite small, finding almost flat $K$-factors in the whole range under analysis.

%

Finally, we want to evaluate the applicability of the HEFT for the computation of the QCD corrections for triple Higgs production.
Formally corresponding to the large top quark mass limit, this approximation completely fails to reproduce the known LO result.
However, it should be more reliable for the computation of the radiative corrections, once the exact LO is used to normalize the latter.
In fact, for the double Higgs production cross section
(where the main contribution comes from the region in which the invariant mass of the di-Higgs final state is larger than the threshold $2m_t$)
it has been shown that the NLO HEFT prediction overestimates the full NLO result by a $14\%$ and $24\%$ for $E_{cm} = 14$ and $100\text{ TeV}$ \cite{Borowka:2016ehy,Borowka:2016ypz} (and with deviations of the same order for the shape of the invariant mass distribution).

Of course, for triple Higgs production the exact NLO result is not available.
However, in order to estimate the level of accuracy of the HEFT we can rely on the approximate NLO results presented in Ref. \cite{Maltoni:2014eza}, where the exact real corrections were included via a reweighting technique.
This approximation, for instance, improves the HEFT result for double Higgs production, overestimating the $14$ and $100\text{ TeV}$ total cross sections by only $4\%$ and $6\%$ respectively.

Using the same setup of Ref. \cite{Maltoni:2014eza} (in particular the same PDF sets and strong coupling), we find that the NLO HEFT result overestimates their approximate NLO prediction by about $7\%$ and $9\%$ for $E_{cm} = 14$ and $100\text{ TeV}$ respectively.
This relatively small deviation, combined with the good level of accuracy shown by the approximation of Ref. \cite{Maltoni:2014eza} for the di-Higgs production cross section, indicates that, within its expected limitations, the (Born-normalized) HEFT can be used to gauge the size of the QCD higher order corrections for triple Higgs boson production.
At NLO, we estimate the uncertainty of the HEFT approach to be of ${\cal O}(20\%)$.

\section{Conclusions}
\label{sec:conc}

In this paper we have computed, for the first time, the QCD corrections for triple Higgs production via gluon fusion predicted by the HEFT.
Within this approach, we have obtained the NLO cross section, and the soft and virtual contributions of the NNLO result.

We have evaluated the numerical impact of the QCD corrections for the LHC and future hadron colliders, both on the total production cross section and on the final state invariant mass distribution.
Corrections were found to be large, with an increase with respect to the LO of ${\cal O}(100\%)$.
The scale uncertainty was substantially reduced, specially when including the NNLO contributions.
We also observed a better convergence of the perturbative series for the central scale choice $\mu_0 = Q/2$, over $\mu_0 = Q$.

While we cannot expect the HEFT to work as accurately as for single Higgs production, we find that it reproduces the approximate NLO results of Ref. \cite{Maltoni:2014eza}, which include the exact real corrections, to better than a $10\%$. Based on that, and on the level of accuracy of the results of Ref. \cite{Maltoni:2014eza} for the double Higgs production cross section, we can roughly estimate the uncertainty of the HEFT prediction to be of ${\cal O}(20\%)$ for triple Higgs production at NLO.

\bibliography{biblio}

\end{document}